\documentclass[aps,prb,onecolumn,11pt,a4paper,citeautoscript,floatfix]{revtex4}

\usepackage{graphicx}
\usepackage{dcolumn}
\usepackage{bm}
\usepackage{color}
\usepackage{amsmath,amssymb,amsfonts}
\usepackage{multirow}
\usepackage{float}

\newcommand{\Na}{Na$_{2-\delta}$Mo$_6$Se$_6$}
\newcommand{\M}{$M_2$Mo$_6$Se$_6$}
\newcommand{\Tl}{Tl$_2$Mo$_6$Se$_6$}
\newcommand{\In}{In$_2$Mo$_6$Se$_6$}

\newcommand{\LiMo}{Li$_{0.9}$Mo$_6$O$_{17}$}

\begin{document}

\title{A disorder-enhanced quasi-one-dimensional superconductor}

\author{A.P. Petrovi\'c$^1$$^,$$^\star$$^,$$^\dagger$, D. Ansermet$^1$$^,$$^\star$, D. Chernyshov$^2$, M. Hoesch$^3$, D. Salloum$^4$$^,$$^5$, P. Gougeon$^4$, M. Potel$^4$, L. Boeri$^6$and C. Panagopoulos$^1$$^,$$^\dagger$} 

\affiliation{{~$^{1}$Division of Physics and Applied Physics, School of Physical and Mathematical Sciences, Nanyang Technological University, 637371 Singapore.}\\
{$^{2}$Swiss-Norwegian Beamline, European Synchrotron Radiation Facility, 6 rue Jules Horowitz, F-38043 Grenoble Cedex, France.}\\
{$^{3}$Diamond Light Source, Harwell Campus, Didcot OX11 0DE, Oxfordshire, United Kingdom.}\\
{$^{4}$Sciences Chimiques, CSM UMR CNRS 6226, Universit\'e de Rennes 1, Avenue du G\'en\'eral Leclerc, 35042 Rennes Cedex, France.}\\
{$^{5}$Faculty of Science III, Lebanese University, PO Box 826, Kobbeh-Tripoli, Lebanon.}\\
{$^{6}$Institute for Theoretical and Computational Physics, TU Graz, Petersgasse 16, 8010 Graz, Austria.}
}


\begin{abstract}
A powerful approach to analysing quantum systems with dimensionality $d>1$ involves adding a weak coupling to an array of one-dimensional (1D) chains.  The resultant quasi-one-dimensional (q1D) systems can exhibit long-range order at low temperature, but are heavily influenced by interactions and disorder due to their large anisotropies.  Real q1D materials are therefore ideal candidates not only to provoke, test and refine theories of strongly correlated matter, but also to search for unusual emergent electronic phases.  Here we report the unprecedented enhancement of a superconducting instability by disorder in single crystals of {\Na}, a q1D superconductor comprising MoSe chains weakly coupled by Na atoms.  We argue that disorder-enhanced Coulomb pair-breaking (which usually destroys superconductivity) may be averted due to a screened long-range Coulomb repulsion intrinsic to disordered q1D materials.  Our results illustrate the capability of disorder to tune and induce new correlated electron physics in low-dimensional materials.  
\end{abstract}

\maketitle

\let\thefootnote\relax\footnotetext{\scriptsize{{$^\star$}These authors contributed equally to this work.}}
\let\thefootnote\relax\footnotetext{\scriptsize{{$^\dagger$}Correspondence and requests for materials should be addressed to A.P.P. (email: appetrovic@ntu.edu.sg) or C.P. (email: christos@ntu.edu.sg)}}

\newpage

\section{Introduction}

Weakly-interacting electrons in a three-dimensional (3D) periodic potential are well-described by Landau-Fermi liquid theory~\cite{Landau1957}, in which the free electrons of a Fermi gas become dressed quasiparticles with renormalised dynamical properties.  Conversely, in the one-dimensional (1D) limit a Tomonaga-Luttinger liquid (TLL) is formed~\cite{Tomonaga1950,Luttinger1963}, where single-particle excitations are replaced by highly correlated collective excitations.  So far, it has proved difficult to interpolate theoretically between these two regimes, either by strengthening electron-electron ($e^\mathrm{-}$-$e^\mathrm{-}$) interactions in 3D, or by incorporating weak transverse coupling into 1D models~\cite{Boies1995,Carr2001}.  The invariable presence of disorder in real materials places further demands on theory, particularly in the description of ordered electronic ground states.  Q1D systems such as nanowire ropes, filamentary networks or single crystals with uniaxial anisotropy therefore represent an opportunity to experimentally probe what theories aspire to model: strongly correlated electrons subject to disorder in a highly anisotropic 3D environment.  

Physical properties of q1D materials may vary considerably with temperature.  TLL theory is expected to be valid at elevated temperatures, since electrons cannot hop coherently perpendicular to the high-symmetry axis and q1D systems behave as decoupled arrays of 1D filaments.  Phase-coherent single-particle hopping can only occur below temperature $T_\mathrm{x} \leq t_\perp$ (where $t_\perp$ is the transverse hopping integral), at which a dimensional crossover to an anisotropic quasi-three-dimensional (q3D) electron liquid is anticipated~\cite{Boies1995,Giamarchi2003}.  The properties of such q3D liquids remain largely unknown, especially the role of electronic correlations in determining the ground state.  At low temperature, a TLL is unstable to either density wave (DW) or superconducting fluctuations, depending on whether the $e^\mathrm{-}$-$e^\mathrm{-}$ interaction is repulsive (due to Coulomb forces) or attractive (from electron-phonon coupling).  Following dimensional crossover, the influence of such interactions in the q3D state is unclear.  As an example, electrical transport in the TLL state of the q1D purple bronze Li$_{0.9}$Mo$_6$O$_{17}$ is dominated by repulsive $e^\mathrm{-}$-$e^\mathrm{-}$ interactions~\cite{Wakeham2011,Chudzinski2012}, yet a superconducting transition occurs below 1.9~K.  

Disorder adds further complication to q1D materials due to its tendency to localise electrons at low temperature.  For dimensionality $d\leq2$, localisation occurs for any non-zero disorder; in contrast, for $d>2$ a critical disorder is required and a mobility edge separates extended from localised states.  The question of whether a mobility edge can form in q1D materials after crossover to a q3D liquid state is open, as is the microscopic nature of the localised phase.  Disorder also renormalises $e^\mathrm{-}$-$e^\mathrm{-}$ interactions, leading to a dynamic amplification of the Coulomb repulsion~\cite{Finkel'stein1987} and a weaker enhancement of phonon-mediated $e^\mathrm{-}$-$e^\mathrm{-}$ attraction, i.e. Cooper pairing~\cite{Ghosal2001,Feigelman2007,Feigel'man2010,Kravtsov2012}.  We therefore anticipate that disorder should strongly suppress superconductivity in q1D materials, unless the Coulomb interaction is unusually weak or screened.  

In this work, we show that the q1D superconductor {\Na} provides a unique environment in which to study the interplay between dimensionality, electronic correlations and disorder.  Although {\Na} is metallic at room temperature, the presence of Na vacancy disorder leads to electron localisation and a divergent resistivity $\rho(T)$ at low temperature, prior to a superconducting transition.  In contrast with all other known superconductors, the onset temperature for superconducting fluctuations $T_{\mathrm{pk}}$ is positively correlated  with the level of disorder.  Normal-state electrical transport measurements also display signatures of an attractive $e^\mathrm{-}$-$e^\mathrm{-}$ interaction, which is consistent with disorder-enhanced superconductivity.  A plausible explanation for these phenomena is an intrinsic screening of the long-range Coulomb repulsion in {\Na}, arising from the high polarisability of disordered q1D materials.  The combination of disorder and q1D crystal symmetry constitutes a new recipe for strongly correlated electron liquids with tunable electronic properties. 

\section{Results}

\textbf{Crystal and electronic structure of {\Na}}\newline

\begin{figure}[hp]
\centering
\includegraphics [trim=0cm 0cm 0cm 0cm, clip=true, width=0.49\columnwidth] {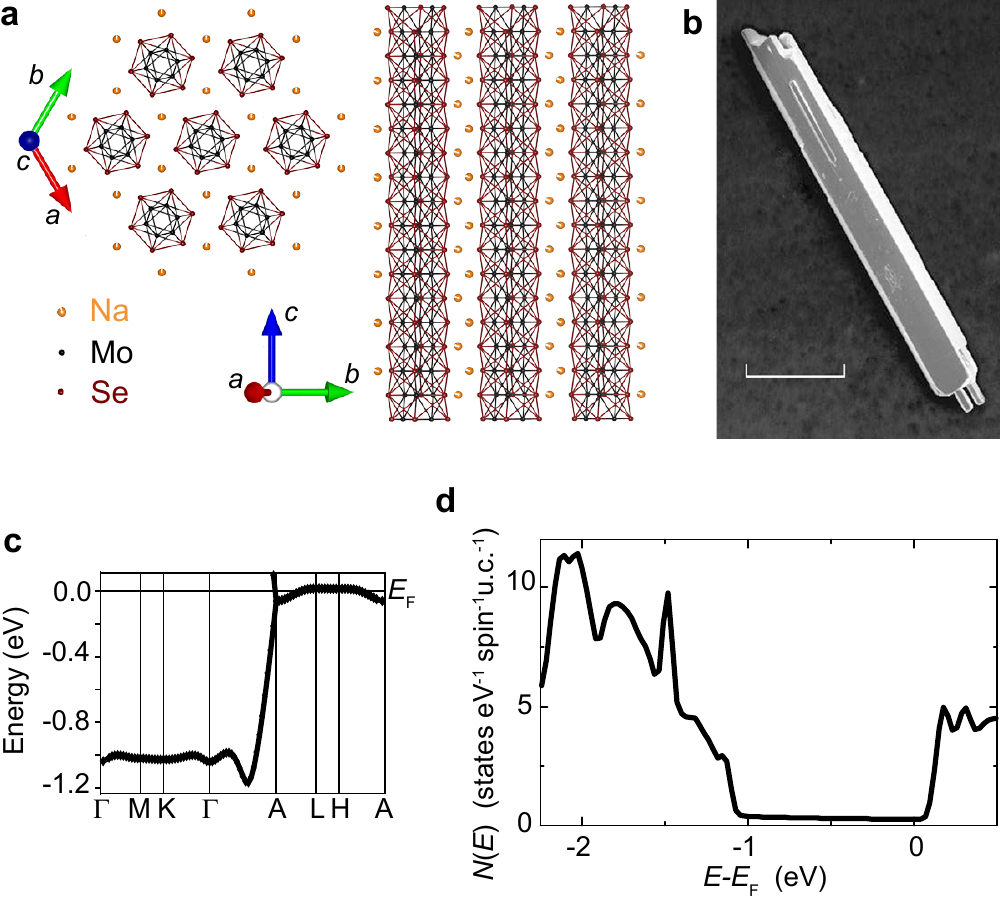}
\caption{\label{Fig1} \textbf{Quasi-one-dimensional crystal and electronic structures in {\Na}}\newline\textbf{a} Hexagonal crystal structure of {\Na}, viewed perpendicular and parallel to the $c$ axis.  From synchrotron X-ray diffraction experiments, we measure the $a$ and $c$ axis lattice parameters to be 8.65~\AA~~and 4.49~\AA~~respectively at 293~K (Supplementary Note II).  \textbf{b} Electron micrograph of a typical {\Na} crystal.  The scale bar corresponds to 300~$\mu$m.  \textbf{c} Calculated energy-momentum dispersion of the conduction band within the hexagonal Brillouin zone, highlighting the large bandwidth and minimal dispersion perpendicular to the chain axis.  \textbf{d} Electronic density of states $N(E)$ around the Fermi level in Na$_2$Mo$_6$Se$_6$.  }  
\end{figure}

{\Na} belongs to the q1D {\M} family~\cite{Potel1980} ($M$ = Group IA alkali metals, Tl, In) which crystallise with hexagonal space group P6$_3/$m.  The structure can be considered as a linear condensation of Mo$_6$Se$_8$ clusters into infinite-length (Mo$_6$Se$_6$)$_\infty$ chains parallel to the hexagonal $c$ axis, weakly coupled by $M$ atoms (Fig.~\ref{Fig1}a).  The q1D nature of these materials is apparent from the needle-like morphology of as-grown crystals (Fig.~\ref{Fig1}b; see Methods for growth details).  \textit{Ab initio} calculations (Supplementary Note I) using density functional theory reveal an electronic structure which is uniquely simple amongst q1D metals.  A single spin-degenerate band of predominant Mo $d_{xz}$ character crosses the Fermi energy $E_\mathrm{F}$ at half-filling (Fig.~\ref{Fig1}c, Supplementary Fig.~1), creating a 1D Fermi surface composed of two sheets lying close to the Brillouin zone boundaries at $\pm\pi/c$ (where $c$ is the $c$-axis lattice parameter).  The warping of these sheets (and hence the coupling between (Mo$_6$Se$_6$)$_\infty$ chains) is controlled by the $M$ cation, yielding values for $t_\perp$ ranging from 230~K ($M$ = Tl) to 30~K ($M$ = Rb) (Supplementary Fig.~2).  In addition to tuning the dimensionality, $M$ also controls the ground state: $M$ = Tl, In are superconductors~\cite{Armici1980a,Petrovic2010}, while $M$ = K, Rb become insulating at low temperature~\cite{Tarascon1984a,Petrovic2010}.  

Within the {\M} family, $M$ = Na is attractive for two reasons.  Firstly, we calculate an intermediate $t_\perp=$~120~K, suggesting that {\Na} lies at the threshold between superconducting and insulating instabilities.  Secondly, the combination of the small Na cation size and a high growth temperature (1750$^\circ$C) result in substantial Na vacancy formation during crystal synthesis.  Since the Na atoms are a charge reservoir for the (Mo$_6$Se$_6$)$_\infty$ chains, these vacancies will reduce $E_\mathrm{F}$ and lead to an incommensurate band filling.  Despite the reduction in carrier density, the density of states $N(E_\mathrm{F})$ remains constant for Na$_{1.5\rightarrow2.1}$ (Fig.~\ref{Fig1}d, Supplementary Note I).  Energy-dispersive X-ray spectrometry on our crystals indicates Na contents from 1.7 to 2, comfortably within this range.  This is confirmed by synchrotron X-ray diffraction (XRD) on three randomly-chosen crystals: structural refinements reveal Na deficiencies of 11$\pm$1\%, 11$\pm$2\% and 13$\pm$4\% (i.e. $\delta$~=~0.22, 0.22, 0.26), but the (Mo$_6$Se$_6$)$_\infty$ chains remain highly ordered.  No deviation from the {\M} structure is observed between 293~K and 20~K, ruling out any lattice distortions such as the Peierls transition, which often afflicts q1D metals.  To probe the Na vacancy distribution, we perform diffuse X-ray scattering experiments on the $\delta$~=~0.26 crystal.  No trace of any Huang scattering (from clustered Na vacancies) or structured diffuse scattering from short-range vacancy ordering is observed (Supplementary Fig.~3, Supplementary Note II).  Na vacancies therefore create an intrinsic, random disorder potential in {\Na} single crystals.  \newline

\textbf{Normal-state electrical transport}

\begin{figure} [hp]
\centering
\includegraphics [width=0.49\columnwidth] {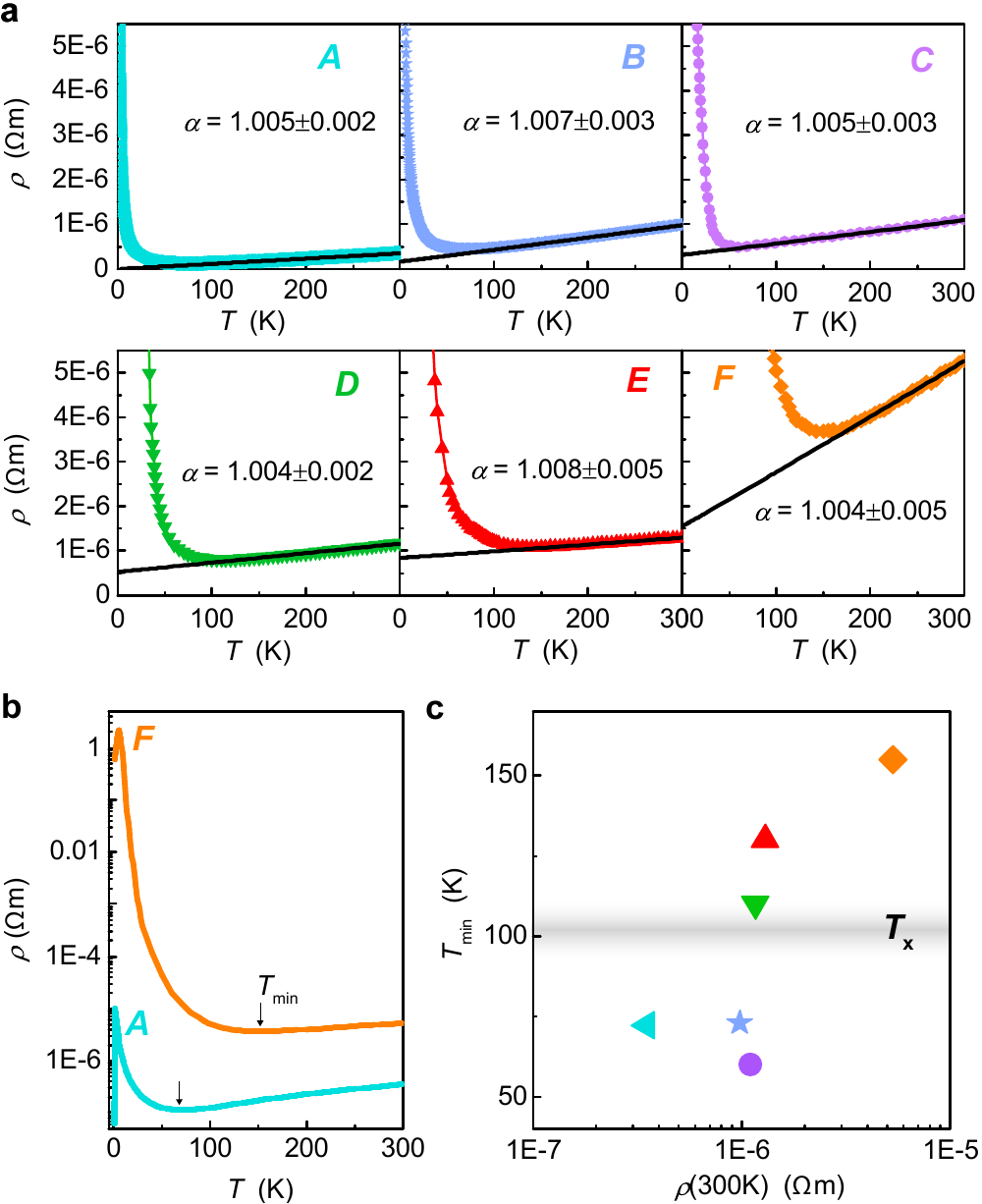}
\caption {\label{Fig2} \textbf{Power-laws and minima in the normal-state resistivity $\rho(T)$}\newline\textbf{a} $\rho(T)$ for crystals $A$-$F$, together with power-law fits $\rho~{\propto}~T^\alpha$ (black lines, fitting range $1.5T_{\mathrm{min}}<T<300$~K).  $T_{\mathrm{min}}$ corresponds to the minimum in $\rho(T)$ for $T>T_{\mathrm{pk}}$.  \textbf{b} $\rho(T)$ plotted on a semilogarithmic scale for crystals $A$ and $F$: $\rho_F\approx10^{5}\rho_A$ as $T \rightarrow T_{\mathrm{pk}}$.  \textbf{c} Evolution of $T_{\mathrm{min}}$ with $\rho(\mathrm{300K})$, which is a measure of the disorder in each crystal.  The horizontal shading indicates the estimated~\cite{Giamarchi2003} single-particle dimensional crossover temperature $T_\mathrm{x}\sim$~104~K, obtained using $T_\mathrm{x} \sim W(t_\perp/W)^{1/(1-\zeta)}$, where $W$ is the conduction bandwidth (Supplementary Note I), $\zeta=(K_\rho+K_\rho^{-1}-2)/8$ and $K_\rho=3/2$.  No anomaly is visible in $\rho(T)$ at $T_\mathrm{x}$, suggesting either that $T_\mathrm{x}$ may be further renormalised due to competing charge instabilities~\cite{Chudzinski2012}, or that signatures of Tomonaga-Luttinger liquid behaviour may persist even for $T<T_\mathrm{x}$~\cite{Giamarchi2003}.  }
\end{figure}

We first examine the electrical transport at high energy for signatures of disorder and one-dimensionality.  The temperature dependence of the resistivity $\rho(T)$ for 6 randomly-selected {\Na} crystals $A$-$F$ is shown in Fig.~\ref{Fig2}a.  $\rho$(300K) increases by more than one order of magnitude from crystal $A$ to $F$ (Fig.~\ref{Fig2}b): such large differences between crystals cannot be attributed to changes in the carrier density due to Na stoichiometry variation and must instead arise from disorder.  Despite the variance in $\rho$(300K), the evolution of $\rho(T)$ is qualitatively similar in all crystals.  Upon cooling, $\rho(T)$ exhibits metallic behaviour before passing through a broad minimum at $T_{\mathrm{min}}$ and diverging at lower temperature.  $T_{\mathrm{min}}$ falls from 150~K to $\sim$~70~K as $\rho$(300K) decreases (Fig.~\ref{Fig2}c), suggesting that the divergence in $\rho(T)$ and the disorder level are linked.  

Upturns or divergence in $\rho(T)$ have been widely reported in q1D materials and variously attributed to localisation~\cite{Sato1987,Narduzzo2007,Enayati-Rad2007,Khim2013,Lu2014}, multiband TLL physics~\cite{DosSantos2008}, DW formation~\cite{DosSantos2007,Xu2009}, incipient density fluctuations~\cite{Petrovic2010} and proximity to Mott instabilities~\cite{Chudzinski2012}.  Differentiating between these mechanisms has proved challenging, in part due to the microscopic similarity between localised electrons and randomly-pinned DWs in 1D.  We briefly remark that the broad minimum in $\rho(T)$ in {\Na} contrasts strongly with the abrupt jumps in $\rho(T)$ for nesting-driven DW materials such as NbSe$_3$~\cite{Ong1977}, while any Mott transition will be suppressed due to the non-stoichiometric Na content.

Instead, a disordered TLL provides a natural explanation for this unusual crossover from metallic to insulating behaviour.  At temperatures $T \gtrsim t_\perp$, power-law behaviour in $\rho(T)$ is a signature of TLL behaviour in a q1D metal.  Fitting $\rho \propto T^\alpha$ in the high-temperature metallic regime of our crystals consistently yields $1<\alpha<1.01$ (Fig.~\ref{Fig2}a).  In a clean half-filled TLL, this would correspond to a Luttinger parameter $K_\rho=(\alpha+3)/4\sim$~1, i.e. non-interacting electrons.  However, disorder renormalises the $e^\mathrm{-}$-$e^\mathrm{-}$ interactions: for a commensurate chain of spinless fermions, $\alpha=2K_\rho-2$ and a critical point separates localised from delocalised ground states at $K_\rho=3/2$~\cite{Giamarchi2003}.  Our experimental values for $\alpha$ therefore indicate that {\Na} lies close to this critical point.  Although the effects of incommensurate band filling on a disordered TLL remain unclear, comparison with clean TLLs suggests that removing electrons reduces $K_\rho$.  For $1<K_\rho<3/2$, $\rho(T)$ is predicted to be metallic at high temperature, before passing through a minimum at $T_{\mathrm{min}}$ (which rises with increasing disorder) and diverging at lower temperature.  These features are consistently reproduced in our data.  

Within the disordered TLL paradigm, our high-temperature transport data indicate that the $e^\mathrm{-}$-$e^\mathrm{-}$ interaction is attractive, i.e. $K_\rho>1$.  This implies that electron-phonon coupling dominates over Coulomb repulsion and suggests that the Coulomb interaction may be intrinsically screened in {\Na}.  A quantitative analysis of the low-temperature divergence in $\rho(T)$ provides further support for the influence of disorder as well as a weak/screened Coulomb repulsion.  We have attempted to fit $\rho(T)$ using a wide variety of resistive mechanisms: gap formation (Arrhenius activation), repulsive TLL power laws, weak and strong localisation (Supplementary Fig.~4, Supplementary Note III).  Among these models, only Mott variable range hopping~\cite{Mott1969} (VRH) consistently provides an accurate description of our data.  VRH describes charge transport by strongly-localised electrons: in a $d$-dimensional material $\rho(T)=\rho_0\exp[(T_0/T)^{\nu}]$, where $T_0$ is the characteristic VRH temperature (which rises as the disorder increases) and $\nu=(1+d)^{-1}$.  Although Mott's original model assumed that hopping occurred via inelastic electron-phonon scattering, VRH has also been predicted to occur via $e^\mathrm{-}$-$e^\mathrm{-}$ interactions in disordered TLLs~\cite{Nattermann2003}.  

\begin{figure*} [hp]
\centering
\includegraphics [width=0.99\columnwidth] {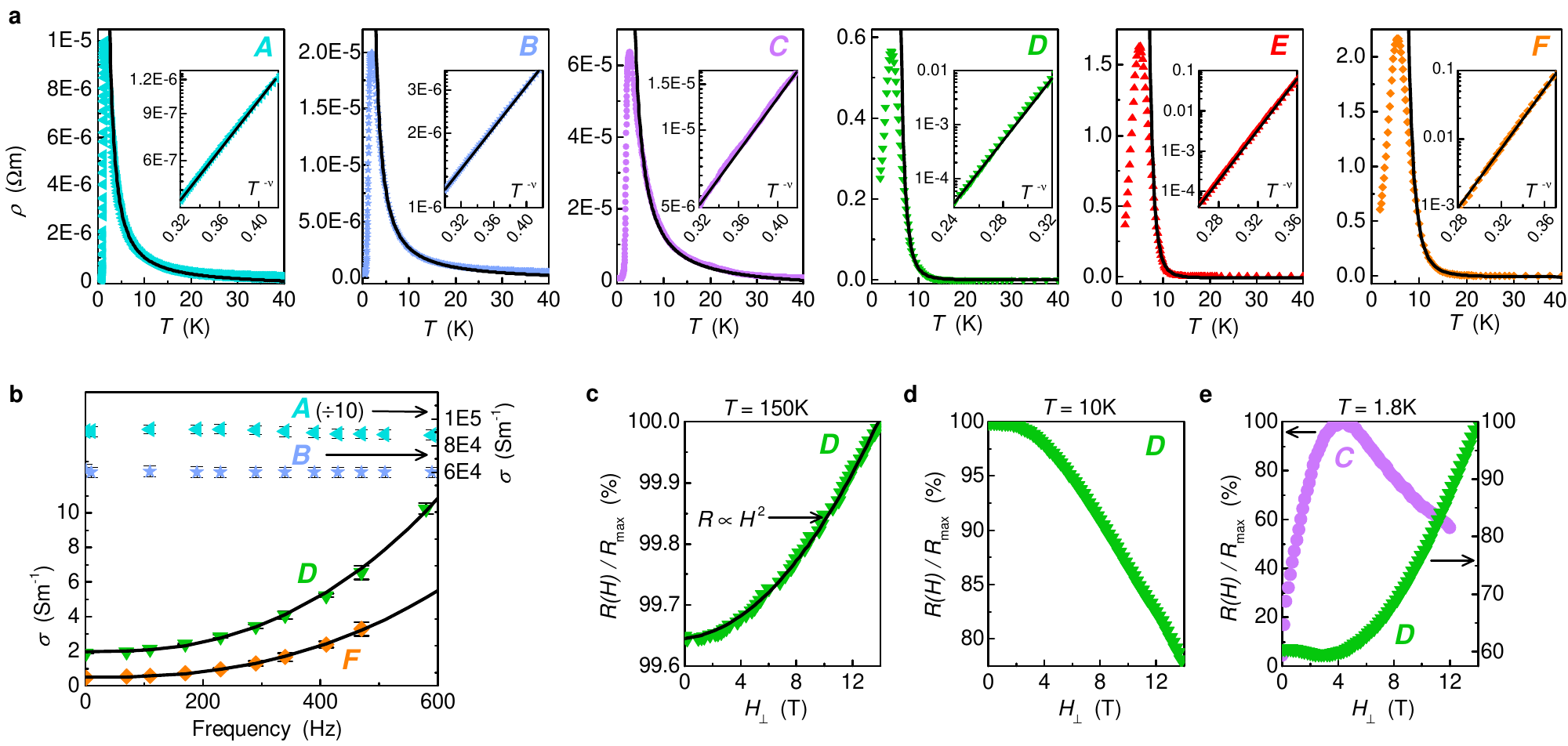} 
\caption {\label{Fig3} \textbf{Influence of electron localisation on the low temperature electrical transport}\newline\textbf{a} Low temperature divergence in the electrical resistivity $\rho(T)$ for six {\Na} crystals $A$-$F$.  Black lines are least-squares fits using a variable range hopping (VRH) model (Supplementary Note III).  $T_0$ (and hence the disorder) rises monotonically from crystal $A$~$\rightarrow$~$F$.  Insets: $\rho(T^{-\nu})$ plotted on a semi-logarithmic scale; straight lines indicate VRH behaviour.  \textbf{b} Frequency-dependent conductivity $\sigma(\omega)$ in crystals $A$,$B$,$D$,$F$ (data-points).  Error bars correspond to the standard deviation in the measured conductivity, i.e. our experimental noise level.  For the highly-disordered crystals $D$,$F$, the black lines illustrate the $\omega^2\log^{(d+2)}(1/\omega)$ trend predicted~\cite{Klein2007} for strongly-localised electrons (using $d=1$). Data are acquired above $T_{\mathrm{pk}}$, at $T=$~4.9, 4.9, 4.6, 6~K for crystals $A$,$B$,$D$,$F$ respectively.  \textbf{c,d,e} Normalised perpendicular magnetoresistance (MR) in crystal $D$ (see Methods for details of the magnetic field orientation).  At 150~K (\textbf{c}) the effects of disorder are weak and $\rho\!\propto\!H^2$ due to the open Fermi surface.  In the VRH regime at 10~K (\textbf{d}), magnetic fields delocalise electrons due to a Zeeman-induced change in the level occupancy~\cite{Fukuyama1979}, leading to a large negative MR.  For $T<T_{\mathrm{pk}}$ (\textbf{e}), the high-field MR is positive as superconductivity is gradually suppressed.  The weak negative MR below $H=3$~T may be a signature of enhanced quasiparticle tunnelling: in a spatially-inhomogeneous superconductor, magnetic field-induced pair-breaking in regions where the superconducting order parameter is weak can increase the quasiparticle density and hence reduce the electrical resistance.  MR data from crystal $C$ are shown for comparison: here the disorder is lower and $H\sim4$~T destroys superconductivity.}
\end{figure*}

Figure~\ref{Fig3}a displays VRH fits for crystals $A$-$F$, while fits to $\rho(T)$ in three further crystals which cracked during subsequent measurements are shown in Supplementary Fig.~5.  All our crystals yield values for $d$ ranging from 1.2 to 1.7 (Supplementary Table I), in good agreement with the $d=$~1.5 predicted for arrays of disordered conducting chains~\cite{Fogler2004}.  Coulomb repulsion in disordered materials opens a soft (quadratic) gap at $E_\mathrm{F}$, leading to VRH transport with $d=1$ regardless of the actual dimensionality.  We consistently observe $d>1$, implying that localised states are present at $E_\mathrm{F}$ and no gap develops in {\Na}.  A small paramagnetic contribution also emerges in the dc magnetisation below $T_{\mathrm{min}}$ and rises non-linearly with $1/T$ (Supplementary Fig.~6).  Similar behaviour has previously been attributed to a progressive crossover from Pauli to Curie paramagnetism due to electron localisation (Supplementary Note IV).  

Although $\rho(T)$ exhibits VRH divergence in all crystals prior to peaking at $T_{\mathrm{pk}}$, a dramatic increase in $\rho(T_{\mathrm{pk}})$ by four orders of magnitude occurs between crystals $C$ and $D$.  This is reminiscent of the rapid rise in resistivity upon crossing the mobility edge in disordered 3D materials.  Our data are therefore suggestive of a crossover to strong localisation and the existence of a critical disorder or ``q1D mobility edge''.  Such behaviour may also originate from proximity to the $K_\rho=3/2$ critical point.  Interestingly, the critical disorder approximately correlates with the experimental condition $T_{\mathrm{min}}\approx T_\mathrm{x}$, where $T_\mathrm{x}$ is the estimated single-particle dimensional crossover temperature (Fig.~\ref{Fig2}c).  This suggests a possible role for dimensional crossover in establishing the mobility edge.  

Further evidence for criticality is seen in the frequency dependence of the conductivity $\sigma(\omega)$ within the divergent $\rho(T)$ regime (Fig.~\ref{Fig3}b).  For crystals with sub-critical disorder, $\sigma(\omega)$ remains constant at low frequency, as expected for a disordered metal.  In contrast, $\sigma(\omega)$ in samples with super-critical disorder rises with frequency, following a $\omega^2\ln^2(1/\omega)$ trend.  This is quantitatively compatible with both the Mott-Berezinskii formula for localised non-interacting electrons in 1D~\cite{Klein2007} and the expected behaviour of a disordered chain of interacting fermions~\cite{Fukuyama1978,Giamarchi2003}.  The strong variation of $\sigma(\omega)$ even at sub-kHz frequencies implies that the localisation length $\xi_\mathrm{L}$ is macroscopic, in contrast with the $\xi_\mathrm{L}\sim(T_0^{1/d}N_{E_\mathrm{F}})^{-1}\lesssim$~100~nm expected from Mott VRH theory~\cite{Shklovskii1984}.  However, it has been predicted that the relevant localisation lengthscale for a weakly-disordered q1D crystal is the Larkin (phase distortion) length, which may be exponentially large~\cite{Fogler2004}.

The evolution of the magnetoresistance (MR) $\rho(H)$ with temperature also supports a localisation scenario.  Above $T_{\mathrm{min}}$, $\rho(H)$ is weakly positive and follows the expected $H^2$ dependence for an open Fermi surface (Fig.~\ref{Fig3}c).  At lower temperature, the divergence in $\rho(T)$ correlates with a crossover to strongly negative MR within the VRH regime (Fig.~\ref{Fig3}d).  The presence of a soft Coulomb gap at $E_\mathrm{F}$ would lead to a positive MR within the VRH regime~\cite{Efros1975}; in contrast, our observed negative MR in {\Na} corresponds to a delocalisation of gapless electronic states~\cite{Fukuyama1979} and provides additional evidence for a screened Coulomb interaction.  The MR switches sign again below $T_{\mathrm{pk}}$ and becomes positive (Fig.~\ref{Fig3}e): as we shall now demonstrate, this is a signature of superconductivity.\newline  

\textbf{Superconducting transitions in {\Na}}

\begin{figure} [hp]
\centering
\includegraphics [width=0.99\columnwidth,clip] {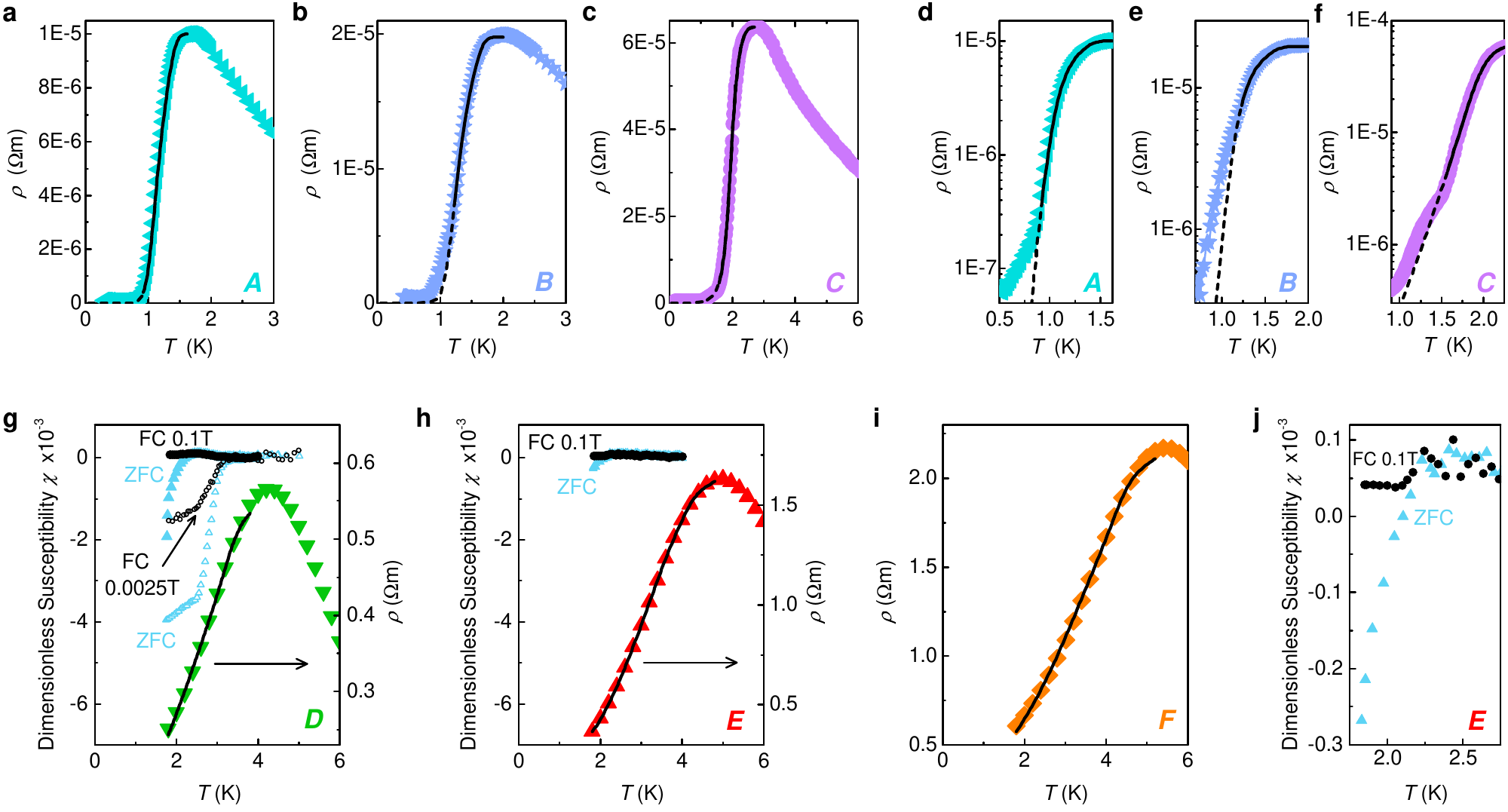} 
\caption{\label{Fig4} \textbf{Resistive and magnetic superconducting transitions in {\Na}}\newline\textbf{a-c} Electrical resistivity $\rho(T<6\mathrm{K})$ for crystals $A-C$.  Coloured points represent experimental data; black lines are fits to a 1D model incorporating thermal and quantum phase slips (Supplementary Note V).  \textbf{d-f} Zoom views of $\rho(T)$ in crystals $A-C$, plotted on a semi-logarithmic scale.  The low temperature limit of our 1D phase slip fits is signalled by a hump in $\rho(T)$, highlighted by the transition from solid to dashed black fit lines: this corresponds to the onset of transverse phase coherence.  In quasi-one-dimensional (q1D) superconductors, such humps form due to finite size or current effects during dimensional crossover~\cite{Ansermet2016}.  \textbf{g-i} $\rho(T<6\mathrm{K})$ for the highly-disordered crystals $D-F$.  Coloured points represent experimental data; black lines are fits to the same 1D phase slip model as in \textbf{a-c}, which accurately reproduces the broad superconducting transitions due to an increased quantum phase slip contribution (Supplementary Note V).  Inhomogeneity and spatial fluctuations of the order parameter are expected to blur the characteristic hump in $\rho(T)$ at dimensional crossover, thus explaining its absence from our data as the disorder rises.  In \textbf{g,h} we also plot zero-field-cooled/field-cooled (ZFC/FC) thermal hysteresis loops displaying the Meissner effect in the magnetic susceptibility $\chi(T)$; \textbf{j} shows a zoom view of the susceptibility in crystal $E$.  Data were acquired with the magnetic field parallel to the crystal $c$ axis and a paramagnetic background has been subtracted.  The small diamagnetic susceptibilities $\left|\chi\right|\ll$~1 are due to emergent pairing inhomogeneity creating isolated superconducting islands~\cite{Feigelman2007}; $\left|\chi\right|$ is further decreased by the large magnetic penetration depth perpendicular to the $c$ axis in q1D crystals.}
\end{figure}

The presence of a superconducting ground state~\cite{Armici1980a,Petrovic2010,Bergk2011} in {\Tl} and {\In} implies that the peak in $\rho(T)$ below 6~K is likely to signify the onset of superconductivity in {\Na}.  Upon cooling crystals $A-C$ in a dilution refrigerator, we uncover a 2-step superconducting transition characteristic of strongly anisotropic q1D superconductors~\cite{Wang2012,Bergk2011,He2015,Ansermet2016} (Fig.~\ref{Fig4}a-c).  Below $T_{\mathrm{pk}}$, superconducting fluctuations initially develop along individual (Mo$_6$Se$_6$)$_\infty$ chains and $\rho(T)$ is well-described by a 1D phase slip model (Supplementary Note V).  Subsequently, a weak hump in $\rho(T)$ emerges (Fig.~\ref{Fig4}d-f) at temperatures ranging from $\sim$~0.95~K (crystal $A$) to $\sim$~1.7~K (crystal $C$).  This hump signifies the onset of transverse phase coherence due to inter-chain coupling.  Cooper pairs can now tunnel between the chains and a Meissner effect is expected to develop, but we are unable to observe this since 1.7~K lies below the operational range of our magnetometer.  Analysis of the current-voltage characteristics indicates that a phase-coherent superconducting ground state is indeed established at low temperature (Supplementary Fig.~7, Supplementary Note VI).  We estimate an anisotropy $\xi_{/\!/}/\xi_\perp=$~6.0 in the coherence length, which is lower than the experimental values for {\Tl} and {\In} (13 and 17 respectively~\cite{Petrovic2010}) in spite of the smaller $t_\perp$ in {\Na} (Supplementary Fig.~2; see Methods for magnetic field orientation details).  This anisotropy is also far smaller than the measured conductivity ratio at 300K: $\sqrt{\sigma_{/\!/}/\sigma_\perp}=$~57.  In comparison, close agreement is obtained between the anisotropies in $\xi_{/\!/,\perp}$ and $\sigma_{/\!/,\perp}$ for Li$_{0.9}$Mo$_6$O$_{17}$~\cite{Mercure2012}, where the effects of disorder are believed to be weak~\cite{Chudzinski2012}.  The disparate anisotropies in {\Na} arise from a strong suppression of $\xi_{/\!/}$, thus illustrating the essential role of disorder in controlling the low temperature properties of {\Na}.

Although superconducting fluctuations are observed regardless of the level of disorder in {\Na}, it is important to identify whether phase-coherent long range order develops in crystals $D$-$F$ which exhibit super-critical disorder.  In Fig.~\ref{Fig4}g-i we demonstrate that $\rho(T)$ in these samples still follows a 1D phase slip model, albeit with a strongly enhanced contribution from quantum phase slips due to the increased disorder~\cite{Altomare2013} (Supplementary Note V).  The fitting parameters for our 1D phase slip analysis are listed in Supplementary Table II.  A weak Meissner effect also develops in the magnetization below $\sim$~3.5~K in crystals $D$ and $E$ (Fig.~\ref{Fig4}g,h,j), but is rapidly suppressed by a magnetic field.  Low transverse phase stiffness is common in q1D superconductors: for example, bulk phase coherence in carbon nanotube arrays is quenched by 2-3~T, yet pairing persists up to 28~T~\cite{Wang2012}.  The superconducting volume fraction corresponding to the magnitude of this Meissner effect is also unusually low: $<$~0.1\%.  Magnetic measurements of the superconducting volume fraction in q1D materials invariably yield values below 100\%, since the magnetic penetration depth $\lambda_{ab}$ normal to the 1D axis can reach several microns~\cite{Petrovic2010} and diamagnetic flux exclusion is incomplete.  For a typical {\Na} crystal of diameter $d\sim$~100~$\mu$m, we estimate that a 0.1\% volume fraction would require $\lambda_{c}\sim$~10~$\mu$m, which seems excessively large.  Conversely, an array of phase-fluctuating 1D superconducting filaments would not generate any Meissner effect at all.  We therefore attribute the unusually small Meissner signal to inhomogeneity in the superconducting order parameter, which is predicted to emerge in the presence of intense disorder~\cite{Ghosal1998,Dubi2007,Feigelman2007,Feigel'man2010}.  In an inhomogeneous superconductor, Meissner screening is achieved via Josephson coupling between isolated superconducting islands~\cite{Beloborodov2007a}.  Within a single super-critically disordered {\Na} crystal, we therefore anticipate the formation of multiple Josephson-coupled networks comprising individual superconducting filaments.  The total magnitude of the diamagnetic screening currents flowing percolatively through each network will be much smaller than that in a homogeneous sample due to the smaller $d/\lambda_{ab}$ ratio, thus diminishing the Meissner effect.\newline

\textbf{Enhancement of superconductivity by disorder}

\begin{figure*} [hp]
\centering
\includegraphics [width=0.99\columnwidth,clip] {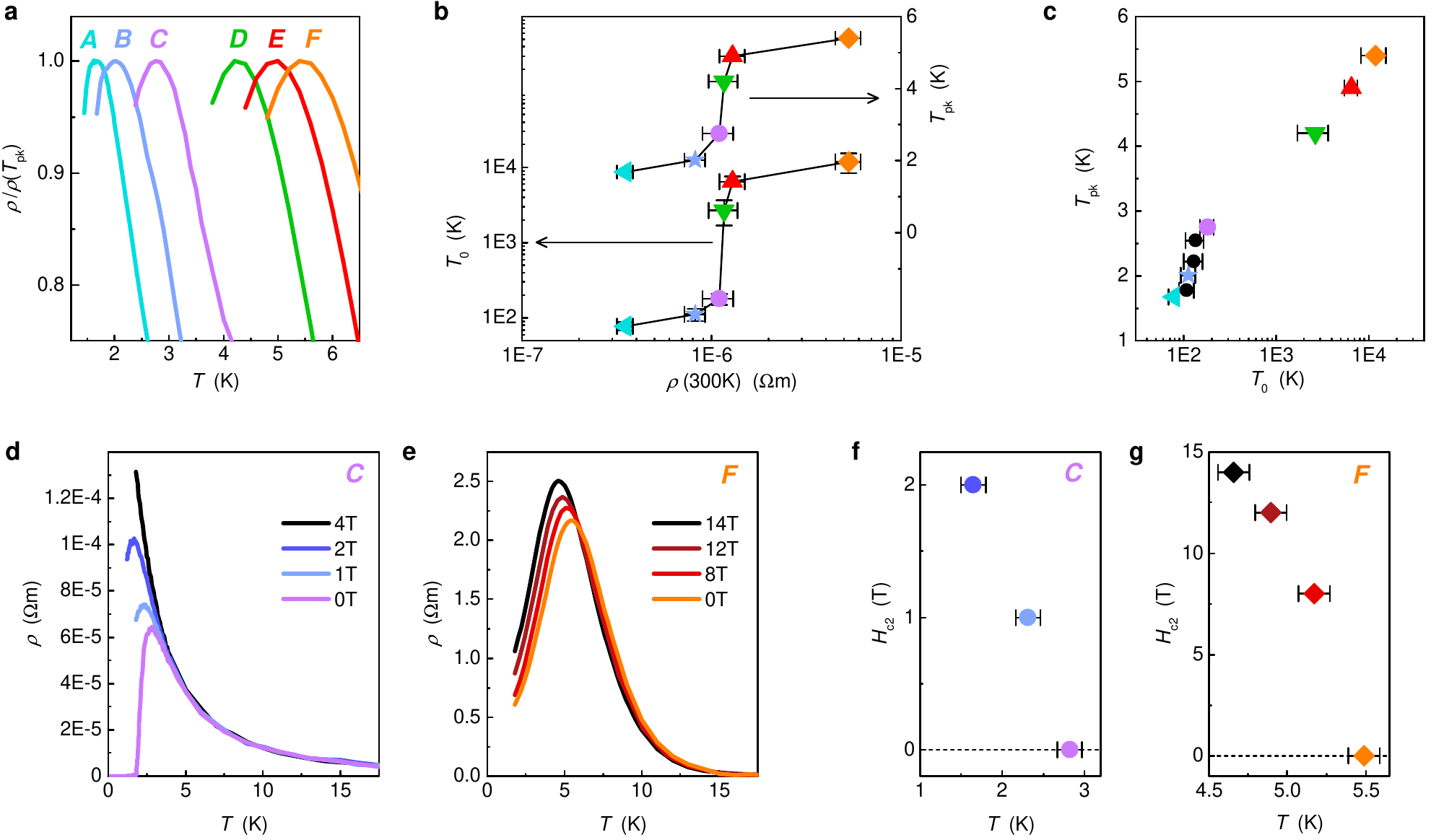} 
\caption{\label{Fig5} \textbf{Disorder controls the divergent electrical resistivity and enhances superconductivity}\newline\textbf{a} Zoom view of the temperature-dependent electrical resistivity $\rho(T)$ at the onset of superconductivity in all crystals, normalised to $\rho(T_{\mathrm{pk}})$.  \textbf{b} Evolution of the characteristic variable range hopping temperature $T_0$ and the superconducting onset temperature $T_{\mathrm{pk}}$ with $\rho(300\mathrm{K})$.  The step at $10^{-6}~\Omega$m corresponds to the critical disorder, i.e. the quasi-one-dimensional mobility edge. Error bars in $\rho(300\mathrm{K})$ are determined from the experimental noise level and our measurement resolution for the crystal dimensions.  The error in $T_0$ corresponds to its standard deviation, obtained from our variable range hopping fitting routine.  \textbf{c} $T_{\mathrm{pk}}$ vs. $T_0$ for each crystal, confirming the positive correlation between superconductivity and disorder.  Data from three additional crystals which broke early during our series of measurements (Supplementary Note III) are also included (black circles).  \textbf{d,e} Suppression of superconductivity with magnetic field $H$ perpendicular to the $c$ axis for crystals $C$ (\textbf{d}) and $F$ (\textbf{e}).  \textbf{f,g} Upper critical field $H_{\mathrm{c}2}(T)$, equivalent to $T_{\mathrm{pk}}(H)$, for crystals $C$ (\textbf{f}) and $F$ (\textbf{g}).  Error bars in $H_{\mathrm{c}2}(T)$ correspond to the error in determining the maximum in $\rho(T,H)$.  $\left.{\partial}H_{\mathrm{c}2}(T)/{\partial}T\right|_{T_{\mathrm{pk}}}$~=~5.1~T~K$^{-1}$ and 24~T~K$^{-1}$ for $C$ and $F$ respectively.}
\end{figure*}

We have established a clear influence of disorder on electrical transport in {\Na} (Figs.~\ref{Fig2},\ref{Fig3}) and demonstrated that the peak in $\rho(T)$ at $T_{\mathrm{pk}}$ corresponds to the onset of superconductivity (Fig.~\ref{Fig4}).  Let us now examine the effects of disorder on the superconducting ground state.  Figure~\ref{Fig5}a illustrates $T_{\mathrm{pk}}$ rising monotonically from crystal $A$ to $F$.  Plotting $T_{\mathrm{pk}}$ as a function of $\rho$(300K) (which is an approximate measure of the static disorder in each crystal), we observe a step-like feature between crystals $C$ and $D$, i.e. at the critical disorder (Fig.~\ref{Fig5}b).  Strikingly, the characteristic VRH temperature $T_0$ which we extract from our $\rho(T)$ fits (Fig.~\ref{Fig3}a) displays an identical dependence on $\rho$(300K).  This implies that disorder controls both the superconducting ground state and the insulating tendency in $\rho(T)$ at low temperature.  The positive correlation between $T_{\mathrm{pk}}$ and $T_0$ (Fig.~\ref{Fig5}c) confirms that the onset temperature for superconducting fluctuations (and hence the pairing energy $\Delta_0$) is enhanced by localisation in {\Na}.  A concomitant increase in the transverse coherence temperature (Supplementary Note VI) implies that some enhancement in the phase stiffness also occurs.

Super-critical disorder furthermore enables superconducting fluctuations to survive in high magnetic fields (Fig.~\ref{Fig5}d-g).  In crystal $C$ (which lies below the q1D mobility edge), superconductivity is completely quenched at all temperatures (i.e. $T_{\mathrm{pk}}\rightarrow0$) by $H=$~4~T (Fig.~\ref{Fig5}d,f).  A giant negative MR reappears for $H>4$~T (Fig.~\ref{Fig3}e), confirming that superconductivity originates from pairing between localised electrons.  In contrast, the peak at $\rho(T_{\mathrm{pk}})$ in the highly-disordered crystal $F$ is strikingly resistant to magnetic fields (Fig.~\ref{Fig5}e,g): at $T$~=~4.6~K, our observed $H_{\mathrm{c2}}=$~14~T which exceeds the weak-coupling Pauli pair-breaking limit $H_\mathrm{P}=$~3~T by a factor $>$~4 (see Supplementary Note VII for a derivation of $H_\mathrm{P}(T)$).  A similar resilience is evident from the positive MR in crystal $D$, which persists up to at least 14~T at 1.8~K (Fig.~\ref{Fig3}e).  Triplet pairing is unlikely to occur in {\Na} (since scattering would rapidly suppress a nodal order parameter) and orbital limiting is also suppressed (since vortices cannot form across phase-incoherent filaments).  Our data therefore suggest that disorder lifts $H_\mathrm{P}$, creating anomalously strong correlations which raise the pairing energy $\Delta_0$~\cite{Ghosal2001,Feigelman2007} above the weak-coupling $1.76k_\mathrm{B}T_{\mathrm{pk}}$.  A direct spectroscopic technique would be required to determine the absolute enhancement of $\Delta_0$, since spin-orbit scattering from the heavy Mo ions will also contribute to raising $H_\mathrm{P}$.

\section{Discussion}

The emergence of a superconducting ground state in {\Na} places further constraints on the origin of the normal-state divergence in $\rho(T)$.  Our electronic structure calculations indicate that the q1D Fermi surface of {\Na} is almost perfectly nested: any incipient electronic DW would therefore gap the entire Fermi surface, creating clear signatures of a gap in $\rho(T)$ and leaving no electrons at $E_\mathrm{F}$ to form a superconducting condensate.  In contrast, our VRH fits and MR data do not support the formation of a DW gap, and a superconducting transition occurs at low temperature.  Electrons must therefore remain at $E_\mathrm{F}$ for all $T>T_{\mathrm{pk}}$, indicating that $\rho(T)$ diverges due to disorder-induced localisation rather than any other insulating instability.  

It has been known since the 1950s that an $s$-wave superconducting order parameter is resilient to disorder~\cite{Abrikosov1959,Anderson1959}, provided that the localisation length $\xi_\mathrm{L}$ remains larger than the coherence length (i.e. the Cooper pair radius).  However, experiments have invariably shown superconductivity to be destroyed by disorder, due to enhanced Coulomb pair-breaking~\cite{Finkel'stein1987}, phase fluctuations~\cite{Kapitulnik1985,Fisher1990,Dubi2007} or emergent spatial inhomogeneity~\cite{Ghosal2001,Bouadim2011}.  In particular, increasing disorder in Li$_{0.9}$Mo$_6$O$_{17}$ (one of the few q1D superconductors extensively studied in the literature) monotonically suppresses superconductivity~\cite{Matsuda1986}.  Therefore, the key question arising from our work is why the onset temperature for superconductivity rises with disorder in {\Na}, in contrast to all other known materials?    

Disorder acts to enhance the matrix element for $e^\mathrm{-}$-$e^\mathrm{-}$ interactions.  This may be explained qualitatively by considering that all conduction electron wavefunctions experience the same disorder-induced potential, developing inhomogeneous multifractal probability densities~\cite{Aoki1983} and hence becoming spatially correlated.  Such enhanced correlations have been predicted to increase the Cooper pairing energy~\cite{Ghosal2001}: in the absence of pair-breaking by long-ranged Coulomb interactions, this will lead to a rise in the superconducting transition temperature~\cite{Feigelman2007,Feigel'man2010,Kravtsov2012,Burmistrov2012,Mayoh2015}.  A proposal to observe this effect in superconducting heterostructures with built-in Coulomb screening~\cite{Burmistrov2012} (by depositing superconducting thin films on substrates with high dielectric constants) has not yet been experimentally realised.  However, our VRH dimensionality $d>1$ (Fig.~\ref{Fig3}a) and negative MR (Fig.~\ref{Fig3}d,e) both point towards a weak or screened Coulomb repulsion, while the power-laws and broad minima in $\rho(T)$ at high temperature (Fig.~\ref{Fig2}a) indicate a Luttinger parameter $K_\rho>1$.  These results all imply that $e^\mathrm{-}$-$e^\mathrm{-}$ interactions in {\Na} are attractive.  (For comparison, $K_\rho\sim0.25$ in {\LiMo} and the $e^\mathrm{-}$-$e^\mathrm{-}$ interaction is repulsive~\cite{Wakeham2011,Chudzinski2012}.)  Phonon-mediated coupling - the Cooper channel - therefore appears to dominate over the Coulomb repulsion in {\Na}, suggesting that the usual disorder-induced Coulomb pair-breaking may be avoided.  Below the q1D mobility edge, our rise in $T_{\mathrm{pk}}$ is quantitatively compatible with a weak multifractal scenario (Supplementary Fig.~8, Supplementary Note VIII), providing a possible explanation for the enhancement of superconductivity which merits further theoretical attention.  

The fact that no experimental examples of q1D materials with attractive $e^\mathrm{-}$-$e^\mathrm{-}$ interactions have yet been reported poses the question why {\Na} should be different.  Although strong electron-phonon coupling is known to play an important role in the physics of molybdenum cluster compounds~\cite{Fischer1978,Petrovic2010}, we propose that the disordered q1D nature of {\Na} may instead play the dominant role, by suppressing the Coulomb repulsion.  In the presence of disorder, a q1D material can be regarded as a parallel array of ``interrupted strands''~\cite{Kuse1971}, i.e. a bundle of finite-length nanowires.  The electric polarisability of metallic nanoparticles is strongly enhanced relative to bulk materials~\cite{Gorkov1965}, although this effect is usually cancelled out by self-depolarisation.  The geometric depolarisation factor vanishes for q1D symmetry, leading to giant dielectric constants $\epsilon$ which rise as the filament length increases~\cite{Rice1972}.  This effect was recently observed in Au nanowires~\cite{Saha2006}, with $\epsilon$ reaching $10^7$.  In {\Na}, we therefore anticipate that the long-range Coulomb repulsion in an individual (Mo$_6$Se$_6$)$_l$ filament ($l<\infty$) will be efficiently screened by neighbouring filaments~\cite{Fogler2004}.  This intrinsic screening provides a natural explanation for attractive $e^\mathrm{-}$-$e^\mathrm{-}$ interactions and suppresses Coulomb pair-breaking in the superconducting phase.  

It has been suggested that impurities can increase the temperature at which transverse phase coherence is established in q1D superconductors~\cite{Efetov1975}.  This effect cannot be responsible for our observed rise in $T_{\mathrm{pk}}$, which corresponds to the onset of 1D superconducting fluctuations on individual (Mo$_6$Se$_6$)$_l$ filaments.  We also point out that the finite-size effects which influence critical temperatures in granular~\cite{Abeles1966} or nanomaterials~\cite{Bose2010} are not relevant in {\Na}: quantum confinement is absent in homogeneously-disordered crystalline superconductors and hence no peaks form in $N(E_\mathrm{F})$.  These mechanisms are discussed in detail in Supplementary Note IX.  

In summary, we have presented experimental evidence for the enhancement of superconductivity by disorder in {\Na}.  The combination of q1D crystal symmetry (and the associated dimensional crossover), disorder and incommensurate band filling in this material poses a challenge to existing 1D/q1D theoretical models.  Although the normal-state electrical resistivity of {\Na} is compatible with theories for disordered 1D systems with attractive electron-electron interactions, we establish several unusual low-temperature transport properties which deserve future attention.  These include a resistivity which diverges following a q1D VRH law for all levels of disorder, the existence of a critical disorder or q1D mobility edge where $T_{\mathrm{min}}\approx T_\mathrm{x}$, and a strongly frequency-dependent conductivity $\sigma(\omega)\sim\omega^2$ in crystals with super-critical disorder.   At temperature $T_{\mathrm{pk}}$, 1D superconducting fluctuations develop, and a phase-coherent ground state is established via coupling between 1D filaments at lower temperature.  As the disorder rises, $T_{\mathrm{pk}}$ increases: in our most-disordered crystals, the survival of superconducting fluctuations in magnetic fields at least four times larger than the Pauli limit suggests that the pairing energy may be unusually large.  

We conclude that deliberately introducing disorder into q1D crystals represents a new path towards engineering correlated electron materials, in remarkable contrast with the conventional blend of strong Coulomb repulsion and a high density of states.  Beyond enhancing superconductivity, the ability to simultaneously modulate band filling, disorder and dimensionality promises a high level of control over emergent order, including density waves and magnetic phases.  More generally, {\Na} and other similar interrupted strand materials may be ideal environments in which to study the evolution of many-body electron localisation beyond the non-interacting Anderson limit.

\section{Methods}
\textbf{Crystal growth and initial characterisation}\newline
A series of {\Na} crystals was grown using a solid-state synthesis procedure.  The precursor materials were MoSe$_2$, InSe, Mo and NaCl, all in powder form.  Before use, the Mo powder was reduced under H$_2$ gas flowing at 1000$^\circ$C for ten hours, in order to eliminate any trace of oxygen.  The MoSe$_2$ was prepared by reacting Se with H$_2$-reduced Mo in a ratio 2:1 inside a purged, evacuated and flame-baked silica tube (with a residual pressure of $\sim$~10$^{-4}$~mbar argon), which was then heated to $\sim$~700$^\circ$C for two days.  InSe was synthesised from elemental In and Se in an evacuated sealed silica tube at 800$^\circ$C for 1 day.  Powder samples of {\Na} were prepared in two steps. Firstly, {\In} was synthesised from a stoichiometric mixture of InSe, MoSe$_2$ and Mo, heated to 1000$^\circ$C in an evacuated sealed silica tube for 36 hours.  Secondly, an ion exchange reaction of {\In} with NaCl was performed at 800$^\circ$C, using a 10\% NaCl excess to ensure total exchange as described in reference~\cite{Potel1984}.  All starting reagents were found to be monophase on the basis of their powder X-ray diffraction patterns, acquired using a D8 Bruker Advance diffractometer equipped with a LynxEye detector (CuK$\alpha_1$ radiation). Furthermore, in order to avoid any contamination by oxygen and moisture, the starting reagents were kept and handled in a purified argon-filled glovebox. 

To synthesise single crystals, a {\Na} powder sample (of mass $\sim$~5~g) was cold-pressed and loaded into a molybdenum crucible, which had previously been outgassed at 1500$^\circ$C for 15~minutes under a dynamic vacuum of $\sim$~10$^{-5}$~mbar. The Mo crucible was subsequently sealed under a low argon pressure using an arc-welding system. The {\Na} powder charge was heated at a rate of 300$^\circ$C/hour up to 1750$^\circ$C, held at this temperature for 3 hours, then cooled at 100$^\circ$C/hour down to 1000$^\circ$C and finally cooled naturally to room temperature within the furnace. Crystals obtained using this procedure have a needle-like shape with length up to 4mm and a hexagonal cross section with typical diameter $\leq$~150~$\mu$m. Initial semi-quantitative microanalyses using a JEOL JSM 6400 scanning electron microscope equipped with an Oxford INCA energy-dispersive-type X-ray (EDX) spectrometer indicated that the Na contents ranged between 1.7 and 2, i.e. up to 15\% deficiency. The Na deficiency results from the high temperatures used during the crystal growth process coupled with the small size of the Na ion: it cannot be accurately controlled within the conditions necessary for crystal growth.

Since {\In} is known to be superconducting below 2.85~K~\cite{Petrovic2010}, it is important to consider the possibility of In contamination in our samples.  The Na/In ion exchange technique used during synthesis is known to be highly efficient~\cite{Potel1984,Tarascon1984} and {\In} decomposes above 1300$^\circ$C, well below our crystal growth temperature (1750$^\circ$C).  This precludes the presence of any superconducting {\In} (or In-rich (In,Na)$_2$Mo$_6$Se$_6$) filaments in our crystals.  Diffuse X-ray scattering measurements accordingly reveal none of the Huang scattering or disk-like Bragg reflections which would be produced by such filaments.  Furthermore, EDX spectrometry is unable to detect any In content in our crystals, while inductively-coupled plasma mass spectrometry indicates a typical In residual of less than 0.01\%, i.e. $<$~0.0002 In atoms per unit cell.  The electronic properties of {\Na} crystals will remain unaffected by such a tiny In residual in solid solution.

\textbf{Electrical transport measurements}\newline
Before all measurements, the as-grown crystal surfaces were briefly cleaned with dilute hydrochloric acid (to remove any residue from the Mo crucible and hence minimise the contact resistance), followed by distilled water, acetone and ethanol.  Four Au contact pads were sputtered onto the upper surface and sides of each crystal using an Al foil mask; 50~$\mu$m Au wires were then glued to these pads using silver-loaded epoxy cured at 70$^\circ$C (Epotek E4110).  Especial care was taken to thoroughly coat each end of the crystal with epoxy, to ensure that the measurement current passed through the entire crystal.  All contacts were verified to be Ohmic at room temperature before and after each series of transport measurements, and at $T=4$~K after cooling.  Typical contact resistances were of the order of 2~$\Omega$ at 300~K.  The transverse conductivity $\sigma_\perp$ was estimated at room temperature using a four-probe technique, with contacts on opposite hexagonal faces of a single crystal.  The temperature dependence of the transverse resistivity $\rho_\perp(T)$ has never been accurately measured in {\M} due to the exceptionally large anisotropies, small crystal diameters and high fragility, even in the least anisotropic {\Tl} which forms the largest crystals~\cite{Armici1980a}.  

Low-frequency four-wire ac conductivity measurements were performed in two separate cryogen-free systems: a variable temperature cryostat and a dilution refrigerator, both of which may be used in conjunction with a superconducting vector magnet.  The ac conductivity was measured using a Keithley 6100 current source, a Stanford SRS850 lock-in amplifier with input impedance 10~M$\Omega$ and (for low resistances, i.e. weakly-disordered samples) a Stanford SR550 preamplifier with input impedance 100~M$\Omega$.  Data from several crystals  were cross-checked using a Quantum Design Physical Property Measurement System (PPMS) with the standard inbuilt ac transport hardware: both methods generate identical, reproducible data.  With the exception of the frequency-dependence studies in Fig.~\ref{Fig3}b, all the transport data which we present in our manuscript are acquired with an ac excitation frequency of 1~Hz, i.e. we are measuring in the dc limit.  At 1~Hz, the phase angle remained zero at all temperatures in all crystals.  Therefore, no extrinsic capacitance effects are present in our data.  

The typical resistance of a weakly disordered crystal lies in the 1-10~$\Omega$ range.  In contrast, the absolute resistances of crystals $D-F$ at $T_{\mathrm{pk}}$ are 41.9~k$\Omega$, 33.7k$\Omega$ and 27.6k$\Omega$ respectively: the crystal diameter increases from $D-F$, thus explaining the rise in resistivity despite a fall in resistance.  These values remain much smaller than our lock-in amplifier input impedance, ruling out any current leakage in highly disordered crystals.  Our measurement current $I_{\mathrm{ac}}=$~10~$\mu$A leads to a maximum power dissipation $<10$~$\mu$W.  This is negligible compared with the $\sim$~2~mW cooling power at 2K on our cryostat cold finger and we may hence rule out any sample heating effects in our data.

We acquire transverse magnetotransport data (Figs.~\ref{Fig3}c-e,~\ref{Fig5}d-g) with the magnetic field perpendicular to both the $c$ axis and the crystal faces, i.e. at 30$^\circ$ to the hexagonal $a$ axis.  Q1D Bechgaard salts and blue/purple bronzes exhibit monoclinic crystal symmetry, and hence strong anisotropies along all three crystallographic axes.  In contrast, {\M} crystallise in a hexagonal lattice: any azimuthal ($\perp c$) anisotropy in {\Na} will therefore reflect this hexagonal symmetry.  In {\Tl}, this anisotropy has been variously reported to be small or entirely absent: it is at least an order of magnitude lower than the polar anisotropy at low temperature~\cite{Lepetit1984}.  Our conclusions regarding the reduced low temperature anisotropy in {\Na} are therefore robust.

In common with most highly one-dimensional materials, {\Na} crystals are extremely fragile, with a tendency to split into a forest of tangled fibres if mishandled.  The crystals therefore exhibit a finite experimental lifetime, with thermal cycling from 2~K to room temperature presenting a particular risk to their structural integrity: this explains why we were unable to obtain complete data-sets in crystals $A$-$F$ (the magnetoresistance $\rho(H)$ at high temperature in crystal $C$ and $\chi(T)$ in Crystal $F$ are missing, for example).

\section{References}

\section{Data Availability}
The authors declare that the data supporting the findings of this study are available within the article and its Supplementary Information files.

\section{Acknowledgements}
We thank Alexei Bosak (Beamline ID28, ESRF Grenoble) for assistance with data collection and processing, and Igor Burmistrov, Vladimir Kravtsov, Tomi Ohtsuki and Vincent Sacksteder IV for stimulating discussions.  The Swiss-Norwegian Beamlines (ESRF Grenoble) are acknowledged for beam time allocation.  This work was supported by the National Research Foundation, Singapore, through Grant NRF-CRP4-2008-04.  

\section{Author contributions}
APP and CP conceived the project; DS, PG and MP grew the crystals; DC performed the XRD measurements with MH and APP; DA carried out the transport experiments; APP and DA analysed the data; LB contributed the electronic structure calculations; APP, DA and CP wrote the paper with input from all the authors; CP supervised the entire project.   

\section{Competing financial interests} The authors declare no competing financial interests.\newline

\newpage

\section{Figure Legends}

\end{document}